\documentclass[12pt]{article}
\usepackage{amssymb,amsmath,amsfonts,latexsym}
\usepackage{graphicx}
\usepackage{amssymb,textcomp}
\usepackage{epsfig}

\DeclareMathOperator{\spn}{span}
\newtheorem{lemma}{Lemma}

\newenvironment{remark}[1][Remark]{\begin{trivlist}
\item[\hskip \labelsep {\bfseries #1}]}{\end{trivlist}}

\begin{document}
\title{Response Operators for Markov Processes in a Finite State Space: Radius of Convergence and Link to the Response Theory for Axiom A Systems}
\author{Valerio Lucarini$^{1,2}$ \\ \small{1. CEN, Institute of Meteorology, University of Hamburg} [\small{\texttt{valerio.lucarini@uni-hamburg.de}}]\\ \small{2. Department of Mathematics and Statistics, University of Reading} \\ }
\vspace{.1em} 
\small{ \date{\today}}
\maketitle
\vspace{-1em} 
\abstract{Using straightforward linear algebra we derive response operators describing the impact of small perturbations to finite state Markov processes. The results can be used for studying empirically constructed - \textit{e.g.} from observations or through coarse graining of model simulations - finite state approximation of statistical mechanical systems. Recent results concerning the convergence of the statistical properties of finite state Markov approximation of the full asymptotic dynamics on the SRB measure in the limit of finer and finer partitions of the phase space are suggestive of some degree of robustness of the obtained results in the case of Axiom A system. Our findings give closed formulas for the linear and nonlinear response theory at all orders of perturbation and provide matrix expressions that can be directly implemented in any coding language, plus providing bounds on the radius of convergence of the perturbative theory. In particular, we relate the convergence of the response theory to the rate of mixing of the unperturbed system. One can use the formulas obtained for finite state Markov processes to recover previous findings obtained on the response of continuous time Axiom A dynamical systems to perturbations, by considering the generator of time evolution for the measure and for the observables. A very basic, low-tech, and computationally cheap analysis of the response of the Lorenz '63 model to perturbations provides rather encouraging results regarding the possibility of using the approximate representation given by finite state Markov processes to compute the system's response.
\section{Introduction}
\subsection{A Brief Summary of Response Theory}
The development of methods for computing the response of a complex system to small perturbations affecting its dynamics is a subject of very active investigation in many fields of science and of technology. Statistical mechanics provide the tool for approaching such a problem through  so-called response theories, which  allow for evaluating the change in the properties of a system through suitably defined operators - response formulas - that factor in  the statistical properties of the unperturbed system and the specific nature of the perturbation one wants to study. 

One can see a response theory as a virtual experimental setting where one has at hand a given system, various measurement instruments, and a knob controlling the value of a parameter, and knows how to relate the position of the knob with the reading of the instruments. In other terms, response theories provide the basis for understanding the outcome of experiments, and, not by chance, physical sciences have been at the forefront of the theoretical investigation in this direction. The monumental contribution by \cite{K57} has provided the basis and the explicit formulas needed for studying the impact of very general perturbations to statistical mechanical systems at equilibrium, as described by the canonical ensemble. The Kubo formulas are extremely useful for studying a large class of problems in \textit{e.g.} transport, optics, and acoustics. A cornerstone of Kubo's theory is the fluctuation-dissipation relation, which enables connecting - within linear approximation - the free fluctuations of the system to its response to perturbations. This property is closely related to the celebrated diffusion law for the brownian motion and has been recently extend to a fully nonlinear case \cite{LC12}. Despite its obvious relevance, Kubo's approach has been criticized  for several reasons:
\begin{itemize}
\item it is not physically consistent in treating the transition from equilibrium to non-equilibrium dynamics, because it studies the impact on equilibrium systems of perturbations that drive them near (but out of) equilibrium, but does not clarify how a new stationary state is reached and maintained, and at the same time it is not suited for studying the response to perturbations  of non-equilibrium systems;
\item it lacks mathematical rigour, as it is not clear which are the systems for which  the response formulas apply, and why it should apply at all.
\end{itemize}
In \cite{R97,R98,R09} it was clarified that it is possible to establish a rigorous response theory for Axiom A \cite{R89} continuous or discrete time dynamical systems. One obtains that the invariant SRB measure is smooth with respect to the  parameter $\epsilon$ that controls the strength of the perturbation changing the dynamics of the system from $\dot{\mathbf{x}}=F(\mathbf{x})$ to $\dot{\mathbf{x}}=\mathbf{F}(\mathbf{x})+\epsilon \mathbf{X}(\mathbf{x})$, in the case of  continuous time evolution, and from $\mathbf{x}_{k+1}=\mathbf{F}(\mathbf{x}_k)$ to $\mathbf{x}_{k+1}=\mathbf{F}(\mathbf{x}_k)+\epsilon \mathbf{X}(\mathbf{x}_k)$, in the discrete case. We continue our discussion in the continuous time case. 

We can introduce the unperturbed evolution operator $S_0^t=\exp(t \mathbf{F}\cdot )$, which moves forward in time any function of phase space $O(\mathbf{x})$ by an interval $t$ according to the unperturbed dynamics, so that  $O(\mathbf{x}(t))=S_0^t O(\mathbf{x}(0))$, and its perturbed counterpart $S_\epsilon^t=\exp(t (\mathbf{F} +\epsilon \mathbf{X})\cdot)$, which instead describes the evolution in the perturbed system. 

We define $\rho_0(\mathrm{d}\mathbf{x})$ and $\rho_\epsilon(\mathrm{d}\mathbf{x})$ the invariant measures of the unperturbed and perturbed states, respectively. In particular, one obtains that the expectation value of sufficiently smooth observables $O(\mathbf{x})$ in the perturbed state can be expressed in the form:
\begin{equation}
[O ]_\epsilon =[ O]_0 + \sum_{j=1}^\infty \epsilon^j \delta\left[O\right]_j,   \label{ruelle1}
\end{equation}
where $[Q]_\epsilon=\int \nu_\epsilon(\mathrm{d}\mathbf{x}) Q(\mathbf{x})$ and $[Q]_0=\int \nu_0(\mathrm{d}\mathbf{x}) Q(\mathbf{x})$, while the various terms of the perturbative expansion can be written as:
\begin{equation}
\delta[O]_j = \int \nu_0(\mathrm{d}\mathbf{x}) \int_0^\infty \mathrm{d} t_1\ldots \int_0^\infty \mathrm{d} t_n \Lambda S_0^{t_1} \ldots S_0^{t_{n-1}} \Lambda S_0^{t_{n}}  O(\mathbf{x})\label{nlruelle},
\end{equation}
where  $\Lambda(\bullet)=\mathbf{X} \cdot \nabla (\bullet)$. In particular, the linear term can be written as:
\begin{equation}
\delta[O]_1 = \int \nu_0(\mathrm{d}\mathbf{x}) \int_0^\infty \mathrm{d} t_1 \Lambda S_0^{t_1}  O(\mathbf{x}),\label{lrruelle}
\end{equation}
All terms $\delta[O]_j$ can be written as an expectation value on the unperturbed measure of a new observable expressed as a functional of  the background vector field $\mathbf{F}$,  of the perturbative vector field $\mathbf{X}$, and of the observable $O$. The somewhat surprising conclusion we draw is that the invariant measure of the system, despite being supported on a strange geometrical set, is differentiable with respect to $\epsilon$. Among the many merits of the Ruelle response theory, one can mention that a) it clarifies the mathematical framework needed for developing a response theory, whose main ingredient, roughly speaking, is the robustness deriving from  having a uniformly hyperbolic dynamics on the attractor supporting an SRB measure; and b) it works seamlessly, in principle, in equilibrium and non equilibrium statistical mechanical systems, reducing to Kubo's formulas when considering the first scenario, \textit{if one assumes that statistical mechanical systems are Axiom A}. Non-trivial implications of the nonequilibrium/equilibrium dichotomy regarding the validity of the fluctuation-dissipation relations are discussed in \cite{R09,LS11,LC12}. 

Of course, at this stage one needs to bridge the gap between mathematical formalism and physical meaningfulness, One manages to bring Ruellle's formalism into the realm of applicability by adopting the \textit{chaotic hypothesis} \cite{GC95,G96}, which basically says that a high-dimensional chaotic physical system can be treated at all practical purposes as if it were Axiom A if we focus on macroscopic observables.  The chaotic hypothesis is the generalisation of the ergodic hypothesis, and provides a firm background for translating the mathematical properties of Axiom A systems into physically meaningful statements. Clearly, the chaotic hypothesis applies far from regimes of metastability and far from critical transitions, where entirely different phenomena appear. The chaotic hypothesis might also be practically problematic in the case one treats multiscale systems featuring many  near-zero Lyapunov exponents; see discussion in  \cite{VL2015}.

Taking the point of view of the chaotic hypothesis, one has that, after transients have died out, nonequilibrium systems reach  a nonequilibrium steady state (NESS) where the phase space is on the average contracting (with the rate of contraction corresponding, broadly speaking, to the entropy production of the system \cite{Gaspard2004}), so that one can associate to the hyperbolic strange attractor supporting the invariant measure a Hausdorff dimension that is lower that the dimensionality of the phase  space and, in general, not integer \cite{R89,G06}. 

The last piece of the puzzle one needs to lay in order to sort out the above-mentioned criticisms to Kubo's theory relies on the physical interpretation of how a perturbed equilibrium system reaches a steady state. A convincing point of view on this relies on emphasizing the role of thermostats, which are large physical systems interacting with the system of interest in such a way to extract the excess of heat generated as result of the energy input due to the  perturbation. Thermostats are also responsible for making it possible the set-up of stationarity in the case of forced and dissipative non equilibrium systems. An extensive treatment of the role of thermostats in equilibrium and nonequilibrium systems in the context of the chaotic hypothesis is given in \cite{G14}. We will not elaborate further on this aspect here.

\subsection{Transfer Operator Approach}
One can point out that  the formulas above describe the impact of  and expressed in terms of expectation values of a generic observable $O$, whereas one might like to derive directly results for the impacts of the perturbations on the invariant measure. 

In \cite{R97,R98,R09} one constructs the response of the system to perturbations by following the changes in the individual trajectories and summing over the possible initial configurations distributed according to the unperturbed invariant measure.  A different point of view on response theory focuses on  studying the properties of the unperturbed and perturbed transfer operators and of their generators (see \cite{B00} for an introduction on these mathematical objects), through the construction of an appropriate framework of suitable (Banach) functional spaces where their actions are well defined, able to carefully treat the fundamental differences between the (smooth) unstable and (singular) stable manifolds of the Axiom A systems \cite{BL07,liverani2008,B08,B14b}.

The evolution of the measure driven by the system $\dot{\mathbf{x}}=\mathbf{F}(\mathbf{x})$ up to time $t\geq 0$ starting from an initial condition at time $t=0$ is described by the Perron-Frobenius transfer $\mathcal{L}^t$  (see, \textit{e.g.}, \cite{B00}), so that $\rho(\mathbf{x},t)= \mathcal{L}^t \rho(\mathbf{x},0)$. We have that the family of $\{\mathcal{L}^t\}_{t\geq 0}$ forms a one-parameter semigroup, such that $\mathcal{L}^{t+s}=\mathcal{L}^t\mathcal{L}^s$ and $\mathcal{L}^0=\mathbf{1}$. The Perron-Frobenius operator $\mathcal{L}^t$ is the adjoint of the evolution operator $S^t=\left(\mathcal{L}^t\right)^\top$, so that $\langle S^t O,\rho\rangle =  \langle O,\mathcal{L}^t\rho\rangle$, where $\langle f,g \rangle$  is the action (computation of the expectation value) of the linear functional $g$ (the probability measure) on the test function $f$ (the observable). We have that $\mathcal{L}^t\nu_0=\nu_0$ $\forall t\geq 0$, meaning that the invariant measure is an eigenvector corresponding to unitary eigenvalue of the Perron-Frobenious operator. 

Assuming strong continuity and boundedness of the semigroup given by $\{\mathcal{L}^t\}_{t\geq 0}$, we can introduce the unperturbed Liouvillian operator $L$, which is the generator of the unperturbed Perron-Frobenius operator $\mathcal{L}^t=\exp(t L)$, and write the Liouville evolution equation for $\rho(\mathbf{x},t)$ as follows \cite{Engel2001}: 
\begin{equation}
\partial_t \rho =-\nabla \cdot \left(\rho \mathbf{F}\right)= L \rho   
\end{equation}
One immediately obtains that $L\nu_0=0$. %In fact, one obtains that there is %The presence of a finite value for $\varepsilon=1-|\lambda_2|$, where $\lambda_2$ is the second largest (in absolute value) eigenvalue of $\mathcal{L}^t$ determines the presence of an exponential decay of correlation 
In general, the spectrum of $L$ is complex and in a strip of finite width including and below the imaginary axis  consists only of isolated eigenvalues
of finite multiplicity  corresponding to the Ruelle-Pollicott resonances, while below such strip one finds the essential spectrum, which is responsible for the continuum of the power spectra of integrable observables. Furthermore, the presence of a unique SRB measure comes from the presence of a simple vanishing eigenvalue, while mixing properties result from the absence of any other eigenvalue along the imaginary axis. The relevance of these properties for constructing a response theory are discussed in great detail in \cite{B08,B14b}. In \cite{chekroun2014} it is argued, using mathematical considerations and examples of geophysical relevance, that the presence of Ruelle-Pollicott resonances having real part close to zero may lead to the presence of rough parameter dependence, as the smoothness of the response if lost. Additionally in \cite{Tantet2015b}, it is shown, along similar lines, that the crisis of a very high-dimensional chaotic attractor near a critical transition - namely, of a climate model in the vicinity of the tipping point responsible for the transition between warm and snowball climate \cite{Hoffman,Pierrehumbert,Luchyst,Lucarini2013a} - can be detected and anticipated by looking at spectrum of the transfer operator.  

We then have that the presence of the $\epsilon$ perturbation to the dynamics changes the Liouville equation as follows: 
\begin{equation}
\partial_t \rho=-\nabla \cdot \left( \rho \mathbf{F}\right) -\epsilon \nabla \cdot \left( \rho \mathbf{X}\right)= L_\epsilon \rho, 
\end{equation}  
so that we can introduce the perturbed Perron-Frobenius operator $\mathcal{L}^t_\epsilon=\exp(tL_\epsilon)$, which pushes forward in time the measure according to the perturbed dynamics: $\rho(\mathbf{x},t)= \mathcal{L}_\epsilon^t \rho(\mathbf{x},0)$. Clearly, $\langle S_\epsilon^t O,\rho\rangle =  \langle O,\mathcal{L}_\epsilon^t\rho\rangle$. Additionally, we have that $\mathcal{L}_\epsilon^t\nu_\epsilon=\rho_\epsilon$ $\forall t\geq 0$ and $L_\epsilon \nu_\epsilon =0$. While this approach is in some sense mathematically more problematic, because it is based on studying a partial differential equation instead of a finite dimensional dynamical system, it seems to provide a more comprehensive set of tools for studying the response of a system and relating it to its unperturbed fluctuations, see, \textit{e.g.}, \cite{BL07}, where Ruelle's formulas are obtained along these lines. See also a comprehensive review given in \cite{B08}, where the applicability of the response theory beyond the case of Axiom A systems is discussed in detail.. %In a finite interval $[-\epsilon_0,\epsilon_0]$ the following recursive formula applies: 
%\begin{equation}
%\frac{\mathrm{d}^n}{\mathrm{d}\epsilon^n}\nu_\epsilon=\int_0^\infty \mathrm{d}t n \frac{\mathrm{d^{n-1}}}{\mathrm{d}\epsilon^{n-1}}\  \nu_n(\mathrm{d}\mathbf{x}),
%\end{equation} 
%nu_\epsilon \mathbf{x})+\sum_{n=1}^\infty \epsilon^n

One needs to emphasise that the transfer operator approach is more natural in all the cases when our interest focuses on studying the properties of the response of an ensemble of trajectories (initialised according to the unperturbed invariant measure) rather than on individual orbits of a system. 

Note that in some applications there is not an obvious separation between the two approaches. Let's take the problem of constructing climate projections through the use of (extremely complex) numerical climate models, which is one of the core activities summarized in the IPCC reports \cite{IPCC13}. Indeed, modelling centers are actively pursuing the preparation of multiple runs starting from an ensemble of initial conditions for a given scenario of forcing in order to estimate more accurately the uncertainties in the projections. Nonetheless, we will not experience an ensemble of realizations of the climatic evolution, but just one. 

\subsection{Computing the Response}\label{responsecomp}
The analysis of high-dimensional complex system in terms of direct numerical simulation and of time series analysis suffers from the (almost) ubiquitous curse of dimensionality, which makes it hard to represent correctly the details of the dynamics because computational complexity explodes with the number of degrees of freedom.  The construction of efficient and accurate algorithms for studying the response of a complex system to perturbations faces serious difficulties. Let's focus now on the linear case. Some previous studies has emphasised the need for treating separately the contributions to the response coming from short and long-time delayed contributions in Eq. \ref{lrruelle}, and have underlined the need for reducing the complexity of the invariant measure by  adding in the background state some stochastic forcing, able to smooth our the singularity of the SRB measure \cite{AM07a,AM07b}. 

A promising way to deal with the actual computation of the scalar product in Eq. \ref{lrruelle} is to use as time-dependent basis the covariant Lyapunov vectors \cite{ER85,GPTCLP07}, which automatically separate the contributions to the response coming from the unstable, neutral, and stable directions. Tis clarifies that the convergence of the formula given in Eq. \ref{lrruelle} comes from the two distinct facts that a) perturbations along the stable directions naturally decay, and b) perturbations along the unstable directions grow in size, but are dominated by the loss of correlation due to mixing.  

Recently, algorithms based upon adjoint methods have shown a good degree of accuracy and seem promising, even if scaling them up to high-dimensional systems has not been attempted yet \cite{EHL04,W13}. A different approach to the problem has been proposed in \cite{L09,LS11,LBHRPW14,RLL14}, where, instead of trying to computing \textit{ab initio} and directly the response given in Eq.  \ref{lrruelle}, the authors construct it \textit{a posteriori}, probing the system with some test forcings and using the formal properties of the theory to be able to predict the response for new patterns of forcings. One can say that by studying the differential response to similar yet differently modulated perturbations, it is possible to derive the overall response properties of the system.    
\subsection{This Paper}\label{thispaper}
Any numerical representation of a continuum system builds upon the need of discretizing the phase space and, in the case of time-continuous system, of time. 
%
%Moreover, in some cases, rather than being interested in studying the details of the (deterministic) dynamics of a given high-dimensional flow, we might be interested in capturing the properties of a given coarse-grained representation. 
%
%
%Moreover, we are able to study the convergence of the perturbation theory by using elementary properties of the stochastic transition matrices, and relate such convergence properties to results on the sensitivity of stochastic matrices to perturbations presented in \cite{Mitro2005,ipsen2011}. 
%
%Coarse graining the dynamics leads unavoidably to introducing a stochastic representation of the process; see \textit{e.g.} the recent contributions  \cite{WL12,WL13} and \cite{CLW15a,CLW15b} for finite and infinite dimensional dynamical system, respectively. 

In this case, we partition the phase space of the system in say $n$ states $\phi_1,\ldots \phi_n$. In many cases, the states are constructed by discretizing the phase space in a grid of boxes, which provide a (Galerkin) basis of orthogonal functions. We then construct an initial ensemble as defined by the occupancy $u_0^1,\ldots,u_0^N$ of each of the $\phi_i$'s, $i=1,\ldots,N$, so that $$u_0^i=\int \mathrm{d}\mathbf{x} \rho(\mathbf{x},0)\mathbf{1}(\phi_i),$$ where $\mathbf{1}(A)$ is the characteristic function in the set $A$, and we want to approximate the evolution of such occupancies change with time, considering discrete time steps $\Delta t$, so that, to a good approximation, $$u^i(k\Delta t)\sim \int \mathrm{d}\mathbf{x} \mathcal{L}^{k\Delta t} \rho(\mathbf{x},0)\mathbf{1}(\phi_i).$$ Moreover, in such a discrete representation, we have that the value of an observable $O$ in the state $\phi_i$ is given by its average 
\begin{equation}
O_j(k\Delta t)=\frac{\int \mathrm{d}\mathbf{x} \rho(\mathbf{x},k\Delta t )\mathbf{1}(\phi_i) O(\mathbf{x})}{\int \mathrm{d}\mathbf{x} \rho(\mathbf{x},k\Delta t)\mathbf{1}(\phi_i)}.\label{expectaton} 
\end{equation}
Let's emphasize that when analyzing virtually any sort of complex system, almost invariably one proposes a natural spatial and temporal cut-off, so that one on not in fact interested in really being able to compute the response of any possible observable defined at any possible spatial and temporal resolution, whereas meso- or macroscopic properties are relevant. Going again to the useful example of climate science, it is commonly regarded as a good and useful question to learn about the change in the surface temperature in response to climate forcing on a spatial scale corresponding to  say a continent or a fraction thereof, and on a temporal scale of say one year. Nobody would find useful nor intelligent to study the surface temperature response over extremely small temporal and spatial scales.  

Empirically, using long numerical integrations and defining the set of finite states $\phi_i$, $i=1,\ldots,N$, we can construct the stochastic matrix $\mathcal{M}_{i,j}$ describing the probability of performing a transition from state $\phi_i$ to state $\phi_j$ in a period of time $\Delta t$. The same operation can in principle be performed using experimental and observational data. A fundamental issue at the core of such procedure is whether for some dynamical systems in the limit of finer and finer partitions covering the phase space (actually, the attractor of the system) with $N\rightarrow \infty$ one reconstructs the actual invariant measure of the original system. See in \cite{Ding2002} a comprehensive discussion of such an issue, the so-called \textit{Ulam conjecture}, and in \cite{Tantet2015a} some extremely promising applications of finite state Markov processes for studying severely reduced representations of complex systems.

Following the idea that the performing the discretization of the phase space amounts to adding a stochastic perturbation of the original dynamical systems, with intensity going to zero with the scale of the actual partitions, and exploiting the fact that the SRB measure can be constructed as zero-noise limit (with measure that is absolutely continuous with respect to Lebesgue)  of the physical measure, in \cite{Froyland1998,Dellnitz1999} it has been proposed that the Ulam conjecture applies in the case of Axiom A systems, which are endowed with an SRB measure.  The convergence in the case of Anosov diffeormorphism has indeed been proved provided one adds some noise of asymptotically vanishing intensity (through stronger than the noise induced by the partition itself) to the underlying dynamics \cite{Blank2002}. Somehow this is not so surprising because by adding noise one introduces a cutoff below which partitions do indeed work. At any practical level, these results suggest that in the case of Axiom A system constructing finite state Markov processes using Ulam partitions can do a pretty good job in simulating the true dynamics, if one consider reasonably well-behaved, smooth observables as test functions. Nonetheless, one has to note that different choices for the partitions can lead to very different rates of convergence \cite{Ding2002}. See also the discussion and the numerical example presented in \cite{Froyland2007}.

Apart from the Ulam method, one can follow a mathematically more elegant but practically much harder way to construct finer and finer partitions. As well known, Axiom A systems possess Markov partitions, \textit{i.e.} well-defined, metric independent, finite resolution representations of the phase space that refine themselves with the dynamics \cite{R89,G14}. Such Markov partitions can be used to construct in the limit the actual SRB measure of the system, and, additionally, following \cite{Froyland1997}, they provide a natural way to build finite Markov chains whose properties converge in the limit to those of the Perron-Frobenius operator of the system. 

Having a response formulas in the finite case has direct relevance for finite Markov chains and for interpreting the results of reduced models. Another good reason to construct a response theory in a finite state space has to do with the fact that the response operators for Axiom A systems introduced by Ruelle can be written as expectation value of certain observables on the unperturbed SRB measure. Therefore, given what said above, one can hope to have convergence of the finite state reconstructed response operators to the corresponding \textit{true} response operator in the limit of infinitely fine partitions of the dynamics. Actually, providing explicit formulas for the response operator for a finite state partition of a system the response operator and taking the limit for (suitably defined) finer and finer partitions could be interpreted as a rigorous way for constructing the actual response on the asymptotic SRB measure.  One needs to note - see discussion in Sects. \ref{convergence} and \ref{observables} - that special attention has to be paid when studying the convergence of such operators. 

%At steady state, we would like to have that, according to our coarse-grained dynamical laws, the occupancies converge to those defined by the invariant measure, so  that $\bar{u}^i\sim\int \nu(\mathrm{d}\mathbf{x})\mathbf{1}(\phi_i)$. 

%Therefore, if we find a good way to describe response theory for a finite state partition of a system and take the results seamlessly to the limit of finer and finer Markov partitions, we can hope, following the results given in, to interpret this as a way to construct that actual response theory on the asymptotic SRB measure. 

In what follows, we present the derivation of the response formulas at all orders of perturbations (as well as the full nonlinear versions)  for finite state spaces of arbitrary size $N$.   All expressions are given in terms of the transitions matrix of the unperturbed system, to its corrections to the perturbation, and of the parameter controlling the strength of the perturbation. The interest we see in the calculations we present below is mostly three-fold:
\begin{itemize}
\item our results are obtained using basic linear algebra operations in finite dimensional spaces, which can used to interpret more complex operators acting on infinite dimensional spaces. It is also possible to use the finite dimensional expressions to derive, \textit{e.g.}, the the actual response operators for continuous time  Axiom A dynamical systems; 
\item we are able to derive an explicit expression for the a lower bound to for the radius of convergence of the perturbative theory, and relate it with the mixing properties of the unperturbed system. We also find a (very tentative) expression for such a lower bound in the case of continuous time case Axiom A dynamical systems;
\item our formulas can be translated into one-line commands in now widely available software tools like \texttt{R}, \texttt{Octave}, or \texttt{MATLAB}$^\circledR$. This might greatly facilitate the actual implementation of response operators. In particular, we can say that our results provide a direct translation of the response theory into a readily implementable algorithms.
\end{itemize}
%The key here is to go from the time-dependent representation of the dynamics to the dynamics of the occupation number of the discrete states, with the result being that in the second picture the convergence of the formulas is automatic. 
%Moreover, we are able to study the convergence of the perturbation theory by using elementary properties of the stochastic transition matrices, and relate such convergence properties to results on the sensitivity of stochastic matrices to perturbations presented in \cite{Mitro2005,ipsen2011}. 

The paper is organised as follows. In Sect. \ref{perturbative}, we introduce some notation and provide basic properties of ergodic finite state Markov chains, which can be taken as mathematical model on its own or as finite precision representation of ergodic (in this case, Axiom A) systems. We also show how it is possible to find an exact expression for the impact of a perturbation on the invariant measure of the Markov process and we study the radius of convergence of the perturbative expansion. In Sec. \ref{observables} we rephrase our results in terms of observables, by constructing straightforward adjoint operators in finite dimensions. In Sect. \ref{continuous} we show how our findings agree with the response theory for continuous time systems when we suitably translate the matrix operations into operators. In Sect. \ref{numerical} we present a simple yet instructive investigation of the response of the Lorenz '63 system \cite{Lorenz:1963} to perturbations using Ulam-like partitions and the formalism developed here. In Sect. \ref{comments} we recapitulate and discuss our results.

\section{Response Operators for Finite-state Markov Processes}\label{perturbative}

Let's consider an ergodic Markov process with a finite number of states defined by the $N$-component vector $\mathbf{u}$. %These states might be chosen according to two different perspectives. On one side, they can be empirically constructed (\textit{e.g.} from observations using mode reduction techniques) finite precision partition of the phase space of a statistical mechanical system. Alternatively, if we are considering an Axiom A, the states can be chosen as an actual  Markov partition with a given level of refinement. 
We consider the infinite Markov chain generated as $\mathbf{u_0}$, $\mathcal{M} \mathbf{u_0}$,$\ldots$ $\mathcal{M}^n \mathbf{u_0}$, $\ldots$ where $\mathbf{u}_0$ is the initial ensemble of states, and $\mathcal{M}_{i,j}\in\mathbb{R}^{N\times N} $ is the stochastic transition matrix determining the probability of reaching the state $i$ at step $n$ if at step $n-1$ we are in the state $j$. The process is taken to be stationary, so that $\mathcal{M}$ does not change with $n$. We remind that $\mathcal{M}$ is such that $\sum_{i=1}^N \mathcal{M}_{i,j}=1$ and $\mathcal{M}_{i,j}\geq 0$ $\forall i,j=1,\ldots,N$. %We consider stationary processes where $\mathcal{M}$ does not depend on $n$ \cite{norris98}. 

%If we are trying to represent a continuous time process, the  steps of the Markov chain describe the dynamics taking place over a finite time interval. If, instead, our focus is coarse-graining a discrete process, each step of the Markov chain can correspond to one or a fixed multiple of steps of the original system.  

The invariant measure is obtained by solving the eigenvalue problem
\begin{equation}
\mathcal{M} \mathbf{u}= \lambda \mathbf{u},
\end{equation}
and selecting the unique solution with eigenvalue $\lambda=1$. The corresponding (column) eigenvector ${\mathbf{u_1}}$\footnote{Most commonly Markov chains are constructed using row vectors; we use column vectors  because we find it easier to perform formal matrix manipulations and because we are closer to the formulation most commonly implemented in scientific software.} is the invariant measure of the system.%, which, if we are treating the problem of a Markov partition of an Axiom A system, gives the finite precision approximation to the SRB measure\footnote{By choosing a more refined Markov partition constructed using the dynamics of the system we will obtain a better approximation to the SRB measure.}. 
We also remind that
\begin{equation}
\lim_{n\rightarrow\infty}\mathcal{M}^n \mathbf{z} = \alpha_1 {\mathbf{u_1}}, \quad \forall \mathbf{z}.
\end{equation}
where $\{\lambda_j,\mathbf{u_j}\}$ $j=1,\ldots,n$ are the pairs of eigenvalues and eigenvectors of $\mathcal{M}$, where $\lambda_1=1$, $|\lambda_j|<1$ if $j>1$, and $\mathbf{z}$ can be expressed as $\mathbf{z}=\sum_{j=1}^n \alpha_j \mathbf{u_j}$.

Our goal is to find a formula for expressing the change in the invariant measure resulting from perturbing the transition matrix $\mathcal{M}\rightarrow \mathcal{M}+\epsilon m$. 
%We redefine $\eta \mathcal{N}=\epsilon m$ where $\epsilon=\eta/||\mathcal{N}||_1$, where the $||\bullet ||_1$ indicates the $L^1$ norm for matrices:
%$$||\mathcal{Q}||_1=\sup_{v\in \mathbb{R}^N,||\mathbf{v}||_1=1}{\frac{||\mathcal{Q}\mathbf{v}||_1}{||\mathbf{v}||_1}},$$
%so that $||m||_1=1$. 

We note that in order to preserve the Markov property of the system, $m$ obeys the following constraint: $\sum_{i=1}^n m_{i,j}=0$, so that $\sum_{i=1}^n\left(\mathcal{M}_{i,j} +\epsilon m_{i,j}\right)=0$. Moreover, an additional constraint on $\epsilon m$  comes from the fact that all elements of  $\mathcal{M} +\epsilon m$ have to be positive. We define
\begin{equation}
\epsilon_+=\max_\epsilon | \forall i,j \in\{1,\ldots,N\}, \mathcal{M}_{i,j} +\epsilon m_{i,j}\geq 0\label{eps+},
\end{equation}
and
\begin{equation}
\epsilon_-=\min_\epsilon | \forall i,j \in\{1,\ldots,N\}, \mathcal{M}_{i,j} +\epsilon m_{i,j}\geq 0\label{eps-};
\end{equation}
clearly, $\epsilon_-\leq 0\leq \epsilon_+$, and the perturbed matrix is a stochastic matrix $\forall \epsilon \in[ \epsilon_-,\epsilon_+]$.  In order to have some \textit{room} for studying the impacts of perturbations, we require that $\epsilon_+-\epsilon_- >0$.   Such conditions show that, for a given $\mathcal{M}$, it makes sense to consider only a specific class of perturbation matrices $m$. Let's provide an example of an ill-chosen $m$: if $\mathcal{M}$ has two zero entries $\mathcal{M}_{i_1,j_1}=\mathcal{M}_{i_2,j_2}=0$ and $m_{i_1,j_1}m_{i_2,j_2}<0$, then we have $\epsilon_-= 0= \epsilon_+$. 
%An additional constraint on $\epsilon$ is described below.
%\section{Response of the Invariant Measure}\label{perturbative}

The new invariant measure is the unique solution to the eigenvalue problem:
\begin{equation}
(\mathcal{M}+\epsilon m) \mathbf{u}= \lambda \mathbf{u},
\end{equation}
with unitary eigenvalue. We define ${\mathbf{v_1}}$  as the invariant measure of the perturbed system. Our goal is to express it as a function of $\mathcal{M}$, $m$, $\epsilon$ and $ {\mathbf{u}}$. This amounts to constructing a response theory. We first present the results of the explicit calculation, and then discuss issues of well-posedness of the problem and convergence of the procedure in Sect. \ref{convergence}. Let's express ${\mathbf{v_1}}={\mathbf{u_1}}+\sum_{j=1}^\infty \epsilon^j {\mathbf{w^{j}}}$, so that we obtain:
\begin{equation}
(\mathcal{M}+\epsilon m) ({\mathbf{u_1}}+\sum_{j=1}^\infty \epsilon^j {\mathbf{w^{j}}})= {\mathbf{u_1}}+\sum_{j=1}^\infty \epsilon^j {\mathbf{w^{j}}},
\end{equation}
Note that the first eigenvalue is \textit{not} changed by the perturbation $\mathcal{M}\rightarrow \mathcal{M} +\epsilon m$, because also $\mathcal{M} +\epsilon m$ is a stochastic matrix. Using  the definition of $\mathbf{u_1}$ we obtain a system of concatenated equations
 \begin{align}
(1-\mathcal{M}) {\mathbf{w^{1}}}&= m{\mathbf{u_1}}\\
(1-\mathcal{M}) {\mathbf{w^{n}}}&=  m{\mathbf{w^{n-1}}}, \quad \forall n\in \mathbf{N}, n>1\label{recur}
\end{align}
We obtain
\begin{align}
{\mathbf{w^{1}}}&= \Psi_1  {\mathbf{u_1}}= (1-\mathcal{M})^{-1}  m {\mathbf{u_1}}\label{recur1}\\
{\mathbf{w^{n}}}&= \Psi_1 {\mathbf{w^{n-1}}}\label{recur2}.
\end{align}
Given the recursive structure, we immediately derive the general formula:
\begin{equation}
{\mathbf{w^{n}}}= \Psi_n  {\mathbf{u_1}} =\Psi_1^n  {\mathbf{u_1}} =  \prod_{j=1}^n\left( \left(1-\mathcal{M}\right)^{-1} m\right) {\mathbf{u_1}}.
\end{equation}
where $\Psi_n=\Psi_1^n$. Concluding, we have that:
\begin{equation}
{\mathbf{v_1}}= {\mathbf{u_1}}+\sum_{n=1}^\infty \epsilon^n {\mathbf{w_n}} ={\mathbf{u_1}}+\sum_{n=1}^\infty \epsilon^n \Psi_1^n {\mathbf{u_1}} = {\mathbf{u_1}} +\sum_{n=1}^\infty \epsilon^n \prod_{j=1}^n\left( \left(1-\mathcal{M}\right)^{-1} m\right){\mathbf{u_1}} \label{perturb}
\end{equation}
which provides the formula we have been looking for. We note that the term responsible for the $n^{th}$ order of perturbation to the measure can be expressed as 
\begin{equation}
\lim_{\epsilon\rightarrow 0 }\frac{1}{n!}\frac{\mathrm{d}^n}{\mathrm{d}\epsilon^n} {\mathbf{v_1}}.
\end{equation}
Using the matrix identity $(1-\mathcal{N})^{-1}= \sum_{k=0}^\infty \mathcal{N}^k$ with $\mathcal{N}=  \epsilon\Psi_1=\epsilon\left(1-\mathcal{M}\right)^{-1} m$, we can also formally express the previous result as:
\begin{equation}
{\mathbf{v_1}}=(1-\epsilon \Psi_1)^{-1} {\mathbf{u_1}}=(1-\epsilon (1-\mathcal{M})^{-1}m)^{-1} {\mathbf{u_1}}.
\end{equation}
Using again the matrix identity $(1-\mathcal{M})^{-1}=\sum_{k=0}^\infty \mathcal{M}^k$, the previous expression can be rewritten as: 
\begin{equation}
{\mathbf{v_1}}=(1-\epsilon \Psi_1)^{-1} {\mathbf{u_1}}=(1-\epsilon \sum_{k=0}^\infty \mathcal{M}^k m)^{-1} {\mathbf{u_1}}.
\end{equation}
or
\begin{equation}
{\mathbf{v_1}}= {\mathbf{u_1}}+\sum_{n=1}^\infty \epsilon^n \prod_{j=1}^n \left(\sum_{k=0}^\infty \mathcal{M}^k m\right) {\mathbf{u_1}} \label{expressfinale}%= {\mathbf{u_1}} +\sum_{n=1}^\infty \epsilon^n \prod_{j=1}^n\left( \left(1-\mathcal{M}\right)^{-1} m\right){\mathbf{u_1}} 
\end{equation}
%is the $n^{th}$ order perturbation operator acting on the invariant measure; note that the product is ordered. Concluding, we have that:
%\begin{equation}
%{\mathbf{v_1}}={\mathbf{u_1}}+\sum_{n=1}^\infty \epsilon^n \Psi_n {\mathbf{u_1}},
%\end{equation}
%which provides the formula we have been looking for. We can also formally express the previous result as:
%\begin{equation}
%{\mathbf{v_1}}=(1-\epsilon \Psi_1)^{-1} {\mathbf{u_1}},
%\end{equation}
%where 
\subsection{Well-posedness and Convergence}\label{convergence}
In the previous equations, we have used somewhat carelessly the expression $(1-\mathcal{M})^{-1}$. Unfortunately, the matrix $1-\mathcal{M}$ is \textit{not} invertible, because all of its columns sum up to zero, or, alternatively, because we know that $1$ is an eigenvalue of $\mathcal{M}$. Nonetheless, the expression makes sense if we apply it to a vector  belonging to $\spn\{\mathbf{u_2},\ldots,\mathbf{u_n}\}$. We now want to prove that:
\begin{lemma}
If $\mathcal{M}$ is a Markov transition matrix $\mathbb{R}^n\rightarrow \mathbb{R}^n$  with eigenvectors $(\mathbf{u_1},\mathbf{u_2},\ldots,\mathbf{u_n})$, and corresponding eigenvalues $(\lambda_1=1,\lambda_2,\ldots,\lambda_n)$, $1>|\lambda_2|\geq\ldots|\lambda_n|$, and $m$ is a matrix matrix $\mathbb{R}^N\rightarrow \mathbb{R}^n$ such that $\sum_{i=1}^n m_{i,j}=0$, then $m\mathbf{z} \in \spn\{\mathbf{u_2},\ldots,\mathbf{u_n}\}$ $\forall \mathbf{z}\in \mathbb{R}^n$.
\end{lemma}
\textbf{Proof}  \quad Let's consider the vector $\mathbf{y}=m\mathbf{z}$. Its $i^{th}$ component can be written as $y_i=\sum_{j=1}^n m_{i,j}z_j$. Since $\sum_{i=1}^n m_{i,j}=0$, we have that $\sum_{i=1}^N z_i= \sum_{i=1}^N \sum_{j=1}^N m_{i,j} z_j=0$.  

Let's now consider the $k^{th}$ eigenvector $\mathbf{u}_k$ of $\mathcal{M}$. We have $\sum_{j=1}^n \mathcal{M}_{i,j} u_{k;j} = \lambda_k u_{k;i}$. Since  $\sum_{i=1}^n \mathcal{M}_{i,j}=1$, taking the sum over the $i$ components of the previous expression, we obtain: $\sum_{i=1}^n \sum_{j=1}^n \mathcal{M}_{i,j} u_{k;j}  = \sum_{j=1}^n u_{k;j} = \lambda_k \sum_{j=1}^n u_{k;j}$. Therefore, either $\lambda_k=1$, or  $\sum_{j=1}^n u_{k;j}=0$. We have that if $k>1$,  $\sum_{j=1}^n u_{k;j}=0$. 

We conclude that $\mathbf{y}=m\mathbf{z} \in \spn\{\mathbf{u_2},\ldots,\mathbf{u_n}\}$ $\forall \mathbf{z}\in \mathbb{R}^n$. 
\begin{remark}
One needs note that finite numerical precision might cause troubles, so that one should be careful in eliminating any component along $\mathbf{u_1}$ at each before applying $\sum_{j=1}^\infty \mathcal{M}^j$. Note that we \textit{must} use $\sum_{j=1}^\infty \mathcal{M}^j$ expression for $(1-\mathcal{M})^{-1}$ in any code, because otherwise any software  would give us automatically a \texttt{NaN} as error message.
\end{remark}
\begin{remark}
We wish to underline another method for avoiding the $\texttt{NaN}$ problem mentioned above. Following \cite{Schweitzer68}, we introduce the fundamental matrix of the Markov chain as $\mathcal{Z}=(1-\mathcal{M}+\mathcal{M}^\infty)^{-1}$, where $\mathcal{M}^\infty$ is the limit matrix whose columns are all equal to $\mathbf{u}$. One can show that $\mathcal{Z}$ exists as the operation of inverse is well defined given the spectral properties of $\mathcal{M}-\mathcal{M}^\infty$ \cite{Froyland1998}. One can show that $\mathcal{M}^\infty m \mathbf{{z}}=0$ $\forall \mathbf{z}\in \mathbb{R}^n$. Therfore, in all the previous Eqs. \ref{recur2}-\ref{expressfinale} we can substitute $(1-\mathcal{M} )^{-1}m=\sum_{j=0}^\infty \mathcal{M}^j m= \mathcal{Z}m=\sum_{j=0}^\infty (\mathcal{M}-\mathcal{M}^\infty)^j m$.
\end{remark}
Let's consider the problem of convergence of the expression in Eq. \ref{perturb}. We want to make sure that the $L^1$ norm of $\sum_{n=1}^\infty \epsilon^n {\mathbf{w_n}}$ does not diverge, and use this to find a bound for the value of $\epsilon$. A simple way to approach this problem is to study the ratio of the $L^1$ norm of two consecutive terms in the previous series. Using Eqs. \ref{recur1}-\ref{recur2}, we have:
\begin{align}\label{domination}
&\frac{\epsilon^n||w_{n}||_1}{\epsilon^{n-1}||w_{n-1}||_1}= \epsilon \frac{|| (1-\mathcal{M})^{-1}m w_{n-1}||_1}{||w_{n-1}||_1} \leq \epsilon \frac{|| (1-\mathcal{M})^{-1}||^*_1 || m w_{n-1}||_1}{||w_{n-1}||_1}\\
&\leq \epsilon || (1-\mathcal{M})^{-1}||^*_1 || m ||_1\label{dom1}\\
&\leq \epsilon  (1-||\mathcal{M}||^*_1)^{-1}|| m ||_1\label{dom2}
\end{align}
where we use the submultiplicative property of the norm and we introduce a modified definition of the $L^1$ norm taking into account that the vector $m\mathbf{v} \in \spn\{\mathbf{u_2},\ldots,\mathbf{u_n}\} \forall \mathbf{v} \in\mathbb{R}^N$: 
$$||\mathcal{Q}||^{*}_1=\sup_{v\in \spn\{\mathbf{u_2},\ldots,\mathbf{u_n}\},||\mathbf{v}||_1=1}{\frac{||\mathcal{Q}\mathbf{v}||_1}{||\mathbf{v}||_1}}.$$
Using expression \ref{dom2}, we have that the perturbative expression converges if 
\begin{equation}
|\epsilon|  (1-||\mathcal{M}||^*_1)^{-1}||m||_1<1 \rightarrow |\epsilon|<   \epsilon_{max}=\frac{1-||\mathcal{M}||^*_1}{||m||_1}\label{epsmaxa},
\end{equation}
%where we have used in the last step that by construction $||m||_1=1$. 
The previous expression  provides an explicit bound for our calculations. We note that $\epsilon_{max}$ is finite because of the restriction imposed in the definition of the norm $|| \bullet ||^*$. Such a bound ensures also the invertibility of $(1-\epsilon \Psi_1)^{-1}.$   
From the previous result, we find the following bound for the first order correction to $$||\epsilon w_1 ||_1 \leq \frac{\epsilon ||m||_1}{1-||\mathcal{M}||^*_1},$$ so that $||m||_1/(1-||\mathcal{M}||^*)$ can be though as a bound to the first order sensitivity of the measure to perturbations. 

Using expression \ref{dom1}, we can derive a more generous bound for $\epsilon$:
\begin{equation}
|\epsilon|<\epsilon^*_{max}=\frac{1}{||m||_1||(1-\mathcal{M})^{-1}||^*_1}\geq \epsilon_{max}\label{epsmax}.
\end{equation}
while $||m||_1||(1-\mathcal{M})^{-1}||^*_1$ provides an additional (stricter) bound to the first order sensitivity. Note that in all the previous expressions we can substitute $||(1-\mathcal{M})^{-1}||_1^*$ with $||\mathcal{Z}||_1$. 

At this point, we wish to refer to previous results (see, \textit{e.g.}, \cite{Mitro2005}) providing bounds for the $L^1$ norm of the difference between the perturbed and unperturbed invariant measure:
\begin{equation}
||v_1-u_1||_1\leq \frac{\epsilon||m||_1}{1-\tau_\mathcal{M}(1)}\label{sensitivity}
\end{equation}
where $\tau_\mathcal{M}(1)$ is the so-called \textit{ergodicity coefficient} \cite{seneta1984} defined as:
$$\tau_\mathcal{M}(1)= \frac{1}{2}\sup_{i,j}||\mathcal{M}(\mathbf{e_i}-\mathbf{e_j})||_1$$ 
with $\mathbf{e_i}$ indicating the unit vector having $1$ at the $i^{th}$ entry and zero elsewhere. We remind that $\tau_1(\mathcal{M})$ is larger than any subdominant eigenvalue of $\mathcal{M}$, and $1/(1-\tau_\mathcal{M}(1))$ can be taken as a definition of conditioning number of $\mathcal{M}$  \cite{ipsen2011}. Clearly if $\tau_\mathcal{M}(1)$ is close to 1, the bound given in Eq. \ref{sensitivity} diverges. Note that $1/(1-\tau_\mathcal{M}(1))$ is the bound to non-perturbative sensitivity mirroring the bound to the perturbative, linearized sensitivity given previously as $1/(1-||\mathcal{M}||_1^*)$. See also  additional results presented in \cite{seneta1993}.

The sensitivity of the unperturbed measure to perturbations given in Eq. \ref{sensitivity} can also be cast in terms $\rho_\mathcal{M}$, the smallest possible value for constant controlling the rate of convergence of iterates $\mathcal{M}\mathbf{e_i}$, $\mathcal{M}^2 \mathbf{e_i}$, $\ldots$, $\mathcal{M}^n \mathbf{e_i}$ to $\mathbf{u_1}$, so that $\forall  n\in\mathbb{N}_+, \forall i\in {1, \ldots N}$ we have that $||\mathcal{M}^n \mathbf{e_i}-\mathbf{u_1}||_1\leq C \rho_\mathcal{M}^n$, $C\geq 1$ \cite{Mitro2005,ipsen2011}. The sensitivity diverges as $\rho_\mathcal{M}$ approaches 1, \textit{i.e.} when the unperturbed matrix has slow properties of convergence. 

While the quantities $||\mathcal{M}||_1^*$, $\tau_{\mathcal{M}}(1)$, and $\rho_\mathcal{M}$ are indeed different, they all point to the fact that if the mixing rate of the unperturbed matrix $\mathcal{M}$ is slow - so that such quantities are close to 1 (so that $||(1-\mathcal{M})^{-1}||_1^*$ and $||\mathcal{Z}||_1$ are very large) -  then the sensitivity of the measure to perturbations is high. See in \cite{chekroun2014} a discussion of the link between slow mixing of a system and the presence of rough parameter dependence in its response to perturbations, with some examples of applications in a geophysical context.

Bringing together the results presented in Eqs. \ref{eps+}-\ref{eps-} and in Eq. \ref{epsmax}, we conclude that Eqs. \ref{perturb}-\ref{expressfinale} provide the exact expression for the invariant measure of the stochastic matrix $\mathcal{M}+\epsilon m$  $\forall \epsilon\in \{[-\epsilon_{max}^*,\epsilon_{max}^*] \cap [\epsilon_-,\epsilon_+]\}$.
%
%Let's define $\lambda_2$ as the second largest (in absolute value) eigenvalue of $\mathcal{M}$. We have from Eq. \ref{recur} that the $n^{th}$ order term in the perturbation series is smaller than the $(n-1)^{th}$ term if $|\epsilon|/(1-|\lambda_2|)=\beta <1$\footnote{Rescaling  the general perturbation matrix $\mathcal{N}$ as $m$ by imposing that $\sigma_1(m)=1$ seems helpful here, in order to avoid prefactors which may change at each order of perturbation.} . 

%Therefore, we must choose $|\epsilon|<(1-|\lambda_2|)$. If the system mixes slowly, so that $1-|\lambda_2|$ is very small, the perturbation theory has a limited domain of applicability. 

%Moreover, we are able to study the convergence of the perturbation theory by using elementary properties of the stochastic transition matrices, and relate such convergence properties to results on the sensitivity of stochastic matrices to perturbations presented in \cite{Mitro2005,ipsen2011}. 

%If the system is non ergodic (the unitary eigenvalue of $\mathcal{M}$ is not unique), the perturbation theory cannot be applied in the present form, because a small perturbation will destroy the degeneracy and change dramatically the structure of the system. 

\section{Response Theory for Observables}\label{observables}
Let's now look at the problem in terms of impact of the perturbation $m$ on the expectation value of observables. Observables live in the dual space of the densities, and, given our convention, they are row vectors. They are approximated as having a constant value within each cell of the chosen partition of the phase space. The expectation value of the observable $\mathbf{\pi}$ with respect to a measure  $\mathbf{w}$ can be written as $\langle \mathbf{\pi},\mathbf{w}\rangle$, where $\langle \bullet , \bullet \rangle$ denotes the scalar product. By definition, we have that $\langle \mathbf{\pi}, A \mathbf{w} \rangle=\langle A^\top \mathbf{\pi}, \mathbf{w}\rangle$, where $A^\top$ indicates the transpose (and adjoint, because we are studying real functions) of $A$.

Let's look at the change in the expectation value of the observable $\pi$ as a result of $\mathcal{M}\rightarrow \mathcal{M}+\epsilon m$. We can write:
\begin{align}
\langle \pi, \mathbf{v_1} \rangle =[\pi]_\epsilon  =&  [\pi]_0 + \sum_{n=1}^\infty \epsilon^n  \delta[\pi]_n \\
						 &=  \langle \pi ,\mathbf{u_1} \rangle + \sum_{n=1}^\infty \epsilon^n \langle \Psi_n^\top  \pi, {\mathbf{u_1}}\rangle\\
						  &=  \langle \pi ,\mathbf{u_1} \rangle + \sum_{n=1}^\infty \epsilon^n \langle \left(\Psi_1^\top\right)^n  \pi, {\mathbf{u_1}}\rangle \label{pertope1},
\end{align}
where  $[\pi]_0=\langle \pi ,\mathbf{u_1} \rangle$ is the expectation value of $\pi$ in the unperturbed system, $[\pi]_\epsilon=\langle \pi ,\mathbf{v_1} \rangle$ is the expectation value of $\pi$ in the perturbed system,  $\delta[\pi]_n$ is the $n^{th}$ order perturbation, which can be expressed as  
\begin{equation}\delta[\pi]_n=\lim_{\epsilon\rightarrow 0 }\frac{1}{n!}\frac{\mathrm{d}^n}{\mathrm{d}\epsilon^n} \langle \pi, \mathbf{v_1} \rangle.\label{sensitivity2}
\end{equation}
Moreover, $\Psi_n^\top$ is the $n^{th}$ order adjoint response operator, acting on the observables, which can be written as:
\begin{equation}
\Psi_n^\top  =(\Psi_1^\top)^n= \prod_{j=1}^n m^\top \left(\sum_{k=1}^\infty \mathcal{M}^k\right)^\top\label{pertope2}.
\end{equation}
We can also wrote Eq. \ref{pertope1} as: 
\begin{align}
\langle \pi, \mathbf{v_1} \rangle& = \langle (1- \epsilon \Psi_1^\top)^{-1} \pi, \mathbf{u_1} \rangle\\
 &= \langle (1- \epsilon m^\top  \left(\sum_{k=1}^\infty \mathcal{M}^k\right)^\top )^{-1} \pi, \mathbf{u_1} \rangle \label{pertope4}\\
 &= \langle (1- \epsilon m^\top(1-  \mathcal{M}^\top)^{-1} )^{-1} \pi, \mathbf{u_1} \rangle \label{pertope5}.
\end{align}
where the last two expressions provide the nonperturbative formulas.
\begin{remark}
Equations \ref{expressfinale} and \ref{pertope1} provide at all orders the response formulas for the discrete Markov process studied here. If we are constructing empirically the discrete phase space, we expect that different choices of the partitions, corresponding to different approximate representations of the full dynamics, will deliver different results in terms of response. Hence, our results can be model dependent, which is acceptable, as we are starting from a subjective choice on the way we approximate the phase space. In fact, one can empirically test the robustness of the obtained results against a set of given criteria by comparing whether the perturbations to a certain set of \textit{relevant} observables weakly depend on the specific partition used. We present a very preliminary (and encouraging) numerical study performed on the Lorenz '63 model \cite{Lorenz:1963} later in Sect. \ref{numerical}.

Moreover, as discussed in Sect. \ref{thispaper}, if we construct finer and finer partitions of for studying the response of systems whose unperturbed dynamics features an SRB invariant measure (most notably in the case of Axiom A systems), and indeed if we follow the self-refining Markov partitions of the dynamics, our results should  converge to the exact response theory built upon the true SRB measure.

One needs to note that Eq. \ref{epsmax} gives an estimate of the largest possible value of $\epsilon$ for \textit{a given partition}, but we are are not sure whether the minimum over all the finer and finer  partitions of $\epsilon_{max}^*$ is positive - this corresponds to imposing the uniform - in $N$ - bound on the norm of $||(1-\mathcal{M})^{-1}||_1^*$ or $||\mathcal{Z}||_1$. 

In \cite{Froyland1998} it is shown that $L^1$ convergence of the finite state measure constructed using the Ulam method to the actual SRB measure is realized when $||\mathcal{Z}||_1$ grows asymptotically not faster than $\log N$, where $N$ is the number of states.  The requirement we seem to have here for applying response theory here is unavoidably stricter because computing the response entails considering the expectation value of not necessarily well behaved observables, constructed through nontrivial operations of differentiation of the actual observables of which we want to study the sensitivity to perturbations, see Eq. \ref{nlruelle} and \cite{R97,R98,R09}. This essential difficulty is exactly what motivates the point of view discussed in \cite{B08,B14}, where a delicate analysis of the relationship between tangent space of the unperturbed dynamics, the perturbation flow, and of the observable allow to set up a robust framework for the response theory. 

Similarly, in our case, making the response theory work at practical level means having/choosing $m$ and $\mathbf{u}$ in such a way that $||(1-\mathcal{M})^{-1}||_1^*$ or $||\mathcal{Z}||_1$ grossly overestimates in terms of norm the effect of applying $(1-\mathcal{M})^{-1}$ or equivalently $\mathcal{Z}$ in, \textit{e.g.}, Eq. \ref{expressfinale}. Additionally, a suitable choice of the observable $\pi$ can help avoiding potential singularities in Eq. \ref{pertope5}. In other terms, response theory can work much more easily once we get rid of or cure pathological cases.

\end{remark}
\section{Towards Continuous Time Dynamical Systems}\label{continuous}
We want to rephrase the previous results in the context of continuous time dynamical systems and derive some formulas previously presented in the literature concerning Axiom A systems. We coonsider a time continuous dynamical system of the form $\dot{\mathbf{x}}=\mathbf{F}(\mathbf{x})$ and study its response to the perturbation $\mathbf{F}(\mathbf{x})\rightarrow \mathbf{F}(\mathbf{x})+\epsilon \mathbf{X}(\mathbf{x})$. Correspondingly, as a result of the perturbation, the original invariant measure $\nu(\mathrm{d}\mathbf{x})$ is changed into $\mu(\mathrm{d}\mathbf{x})$. The Liouville equation describing the evolution of a given initial density of states $\rho(\mathbf{x})$ for the unperturbed system can be written as
\begin{equation}
\partial_t \rho(\mathbf{x},t) =-\nabla\cdot \left(\mathbf{F}(\mathbf{x})\rho(\mathbf{x},t)\right);
\end{equation}
considering two instants of time separated by a small time interval $\mathrm{d}t$, we have:
\begin{align}
&\rho(\mathbf{x},t+\mathrm{d}t) =\rho(\mathbf{x},t) -\mathrm{d}t \nabla\cdot \left(\mathbf{F}(\mathbf{x})\rho(\mathbf{x},t)\right)= \mathcal{M} \rho(\mathbf{x},t)\nonumber\\
&\mathcal{M}=1+\mathrm{d}t\mathcal{F} \quad \mathcal{F}= -\nabla\cdot \left(\mathbf{F}(\mathbf{x})\bullet\right)\label{Mcont}.
%\mathcal{M}=\mathbf{1}-\mathrm{d}t\mathcal{F}=\mathbf{1}-\mathrm{d}t \nabla\cdot \left(\mathbf{F}(\mathbf{x})\bullet\right)\label{Mcont}.
\end{align}
We understand that $\mathcal{M}$ is in this context the unperturbed Perron-Frobenius operator $\mathcal{L}_\epsilon^{\mathrm{d}t}$ pushing forward the measure $\rho$ from $t$ to $t+\mathrm{d}t$. When looking at the perturbed flow we have:
\begin{align}
\rho(\mathbf{x},t+\mathrm{d}t) &=\rho(\mathbf{x},t) -\mathrm{d}t \nabla\cdot \left(\mathbf{F}(\mathbf{x})\rho(\mathbf{x},t)\right)-\mathrm{d}t \epsilon \nabla\cdot \left(\mathbf{X}(\mathbf{x})\rho(\mathbf{x},t)\right)\nonumber \\
&=(\mathcal{M}+\epsilon m) \rho(\mathbf{x},t), 
\label{mcont}.
\end{align}
where
\begin{equation}
%\mathcal{M}=1+\mathrm{d}t\mathcal{F} \quad \mathcal{F}= -\nabla\cdot \left(\mathbf{F}(\mathbf{x})\bullet\right)\quad 
m=\mathrm{d}t\mathcal{X} \quad \mathcal{X} =- \nabla\cdot \left(\mathbf{X}(\mathbf{x})\bullet\right)\label{mcont2}
\end{equation}

In this case, starting from Eq. \ref{domination}, and considering that no normalization is applied to the perturbation operator, it is possible to propose a definition of $\epsilon^*_{max}$ for the continuous time dynamics taking inspiration from Eq. \ref{epsmax}:
\begin{equation}
\epsilon^*_{max}=\frac{1}{||\mathcal{X}||_\mathcal{B}||\mathcal{F}^{-1}||^*_\mathcal{B}},
\end{equation}
such that the perturbative expansion converges if $\epsilon\leq \epsilon^*_{max}$,  where $|| \bullet ||_\mathcal{B}$ describes the norm of the operator in the appropriate Banach space $\mathcal{B}$ it belongs to, while $|| \bullet ||^*_\mathcal{B}$ is such that the computation of the norm excludes the SRB measure. Note that $\epsilon^*_{max}$ is finite if both  $||\mathcal{X}||_\mathcal{B}||$ and $||\mathcal{F}^{-1}||^*_\mathcal{B}$ are finite. This expression is admittedly tentative. As mentioned before, the problem of selecting appropriate functional spaces for constructing the response theory for Axiom A systems along the lines of studying the perturbations to the transfer operator requires a careful construction of suitable Banach spaces and of the related metrics \cite{BL07,B08,B14b} and is beyond the scope of this paper.\footnote{Following \cite{liverani2006}, one might tentatively consider the norms of the operator acting between the Banach spaces $\mathcal{B}_{2,q}$ and $\mathcal{B}_{1,q+1}$.} 

\subsection{Linear response}
We now want to derive the Ruelle response formulas for computing the linear correction to the invariant measure resulting from the perturbation. We write
\begin{equation}
\nu_\epsilon(\mathrm{d}\mathbf{x})=\nu_0(\mathrm{d}\mathbf{x})+\sum_{n=1}^\infty \epsilon^n \nu_n(\mathrm{d}\mathbf{x}),
\end{equation} 
where $n$ indicates the order of perturbation.
Let's first go back to the first order term in Eq. \ref{recur1}:
\begin{equation}
\epsilon{\mathbf{w^{1}}}=\epsilon \Psi_1 \mathbf{u_1} =  \left(\sum_{k=1}^\infty \mathcal{M}^k \right) m {\mathbf{u_1}}.
\end{equation}
Each term of the form $\mathcal{M}^k$ pushes forward up to time $t_k=k\times \mathrm{d}t$ what is positioned to its right. Summing over $k$ in, in fact, amounts to looking forward in time. If we insert the definition of $m$ given above, we get the integrating factor $\mathrm{d}t$, so that we obtain the following expression:
\begin{equation}
\nu_1(\mathrm{d}\mathbf{x})= - \int_0^\infty \mathrm{d} t  \nabla \cdot (\mathbf{X}(\mathbf{x}(t)) \nu(\mathrm{d}\mathbf{x})),
\end{equation}
where the evolution takes place according to the unperturbed system, and we have used the invariance of $\nu(\mathrm{d}\mathbf{x})$ with respect to such an evolution law. 

By going into the dual space of the observables, we have that the change in the value of an observable $O(\mathbf{x})$ from time $t$ to time $t+\mathrm{d}t$ in the unperturbed system can be written as:
\begin{equation}
\frac{\mathrm{d}}{\mathrm{d}t} O(\mathbf{x}(t))= \mathbf{F}(\mathbf{x})\cdot \nabla O(\mathbf{x}(t)), 
\end{equation}
so that 
\begin{equation}
O(\mathbf{x}(t+\mathrm{d}t))=O(\mathbf{x}(t)) +\mathrm{d}t \mathbf{F}(\mathbf{x})\cdot \nabla O(\mathbf{x}(t))=\mathcal{M}^\top O(\mathbf{x}(t)).
\end{equation}
where the operator $\mathcal{M}^\top=1+\mathrm{d}t \mathcal{F}^\top = \mathbf{1}+\mathrm{d}t \mathbf{F}(\mathbf{x})\cdot \nabla (\bullet)$. Along the same lines, one derives that the perturbation operator $m^\top$ acting on the observable can be written as $m^\top=\mathrm{d}t\mathcal{X}^\top=\mathrm{d}t\mathbf{X}(\mathbf{x})\cdot \nabla (\bullet)$. Furthermore, we introduce the following expansion for the expectation value of $O(\mathbf{x})$:
\begin{equation}
[O]_\epsilon=[O]_0+\sum_{n=1}^\infty \epsilon^ n\delta[O]_n,
\end{equation}
where $[O]_\epsilon$ is the expectation value in the perturbed system, $[O]_0$ is the unperturbed expectation value, and the corrections are included in the summation.

Applying this expression to the first order term in Eq. \ref{pertope1}-\ref{pertope2}:
\begin{equation}
\delta[\pi]_1 = \epsilon \langle \Psi_1^\top \pi, \mathbf{u}_1\rangle  = \epsilon \langle \left( m^\top \left(\sum_{k=1}^\infty \mathcal{M}^k\right)^\top \right) \pi, \mathbf{u}_1\rangle \label{pertope3}.
\end{equation}
we get:
\begin{equation}
\delta[O]_1 = \epsilon \int \nu(\mathrm{d}\mathbf{x}) \int_0^\infty \mathrm{d} t \mathbf{X}(\mathbf{x}) \cdot \nabla O(\mathbf{x}(t))= \epsilon \int \nu(\mathrm{d}\mathbf{x}) \int_0^\infty \mathrm{d} t \Lambda S^t_0 O(\mathbf{x})\label{linearruelle}
\end{equation}
which is exactly the original version of Ruelle's linear response formula given in Eq. \ref{lrruelle}. 

One needs to note that what in Ruelle's formulation is causality (time integration in the response starts from 0), in the context of the Markov matrices formalism followed here comes from the algebraic expansion of $(1-\mathcal{M})^{-1}$. The issues of convergence mentioned in the original paper by Ruelle can be translated in the rate of mixing of the system as determined by the properties of $\mathcal{M}$ discussed in Sect. \ref{convergence}. 
\subsection{Higher order terms}
We can repeat the same construction to derive the higher order perturbation terms in the case of the continuous time dynamical systems. Inserting in Eqs. \ref{recur1}-\ref{recur2} the expression \ref{Mcont} for $\mathcal{M}$ and expression \ref{mcont} for $m$, we obtain for the second order the following expression for the perturbation to the invariant density:
\begin{equation}
\nu_2(\mathrm{d}\mathbf{x})=\epsilon^2 \int_0^\infty \mathrm{d}t_1\int_0^\infty\mathrm{d}t_ 2  \nabla \cdot \left(\mathbf{X}(\mathbf{x}(t_1) \nabla_{\mathbf{x}(t_1)} \cdot \left(\mathbf{X}(\mathbf{x}(t_1+t_2)\right)\right)\nu(\mathrm{d}\mathbf{x}),
\end{equation}
while the expression for the $n^{th}$ order correction reads like
\begin{equation}
\nu_n(\mathrm{d}\mathbf{x})=(-1)^n\epsilon^n \int_0^\infty \mathrm{d}t_1\ldots \int_0^\infty\mathrm{d}t_ n  \nabla \cdot \left(\mathbf{X}(\mathbf{x}(t_1) \ldots \nabla_{\mathbf{x}(t_1+\ldots t_{n-1})}  \cdot \left(\mathbf{X}(\mathbf{x}(t_1+\ldots t_{n})\right)\right)\nu(\mathrm{d}\mathbf{x}),
\end{equation}
Considering the adjoint problem and computing the higher order corrections to the expectation value of the observable $O$, we derive the general response formula proposed by Ruelle
\begin{equation}
\delta[O]_n = \epsilon^n \int \nu(\mathrm{d}\mathbf{x}) \int_0^\infty \mathrm{d} t_1\ldots \int_0^\infty \mathrm{d} t_n \Lambda S^{t_1} \ldots S^{t_{n-1}}  \Lambda S^{t_{n}}  O(\mathbf{x})\label{nonlinearruelle},
\end{equation}
as reported in Eq. \ref{nlruelle}. 

\section{A very basic numerical experiment}\label{numerical}
In order to make a (very) preliminary assessment of the potential of some of the ideas presented in this paper, we have focused on investigating some properties of the celebrated Lorenz '63 system \cite{Lorenz:1963}: 
\begin{eqnarray}\label{lorenz}
\dot{x}=&\sigma(y-x)\nonumber\\
\dot{y}=&x(\rho-z)-y\\
\dot{z}=&xy-\beta z\nonumber
\end{eqnarray}
where we have chosen the standard value for the parameters $\sigma =10$, $\rho=28$, and $\beta=8/3$. We remark that such a system is not an Axiom A, but instead a singular hyperbolic system \cite{Bonatti2005}, which possesses a chaotic attractor and an invariant SRB measure \cite{Tucker}. In a previous publication \cite{L09}, we have performed an  analysis of the linear and nonlinear response of the Lorenz '63 to perturbations, extending a previous investigation by Reick \cite{reick02}, which makes us confident that response theory can be safely applied at all practical purposes also in this case. We consider the special case of time-indepedent perturbations to the dynamics resulting from substituting $\rho \rightarrow \rho+\epsilon$ in Eq. \ref{lorenz}, so that the perturbation flow can be written as $\epsilon \mathbf{X}(\mathbf{x})=[0\quad  \epsilon x\quad  0]^\top$. 

We have then identified a 3-dimensional box $\mathcal{B}$ containing the attractor, defined as $\mathcal{B}=\{(x,y,z)\in\mathcal{R}^3 |x\in[-20,20],\quad y\in[-30,30],\quad z\in[-0,50]\}$, and subdivided it, \'a la Ulam, in smaller boxes of size using a regularly spaced cartesian grid. We have considered partitions obtained using small boxes with linear dimension given by $dx=2 \times j$, $dy=3\times j$, and $dz=2.5\times j$, along the three directions, with $j=1,2,4$, see Fig. \ref{lorenzattr}. This amounts to partitioning $\mathcal{B}$ into $8000/j^3$ smaller boxes. Note that our construction delivers a much lower resolution with respect to what used in, \textit{e.g.}, \cite{Froyland2009}.

We run the model with standard values of the parameters choosing as initial condition $[1\quad 1 \quad 1]^\top$ (in fact, given the global attractivity and ergodicity of the Lorenz attractor, any initial condition can be chosen), and, after discarding a transient of 1000 time units, which brings us safely into the asymptotic regime, we run the model for $50000$ time units with a simple  Runge-Kutta $4^{th}$ order adaptive scheme and obtain the output with time step of 0.001 time units. This takes less than 10 minutes in a today's commercial laptop with standard specifics using \texttt{MATLAB}$^\circledR$. We present results at such a low level of sophistication in order to clarify that the appracch proposed here is rather robust and of relatively simple implementation.

\begin{figure}[ht]
\centering
\includegraphics[width=140mm]{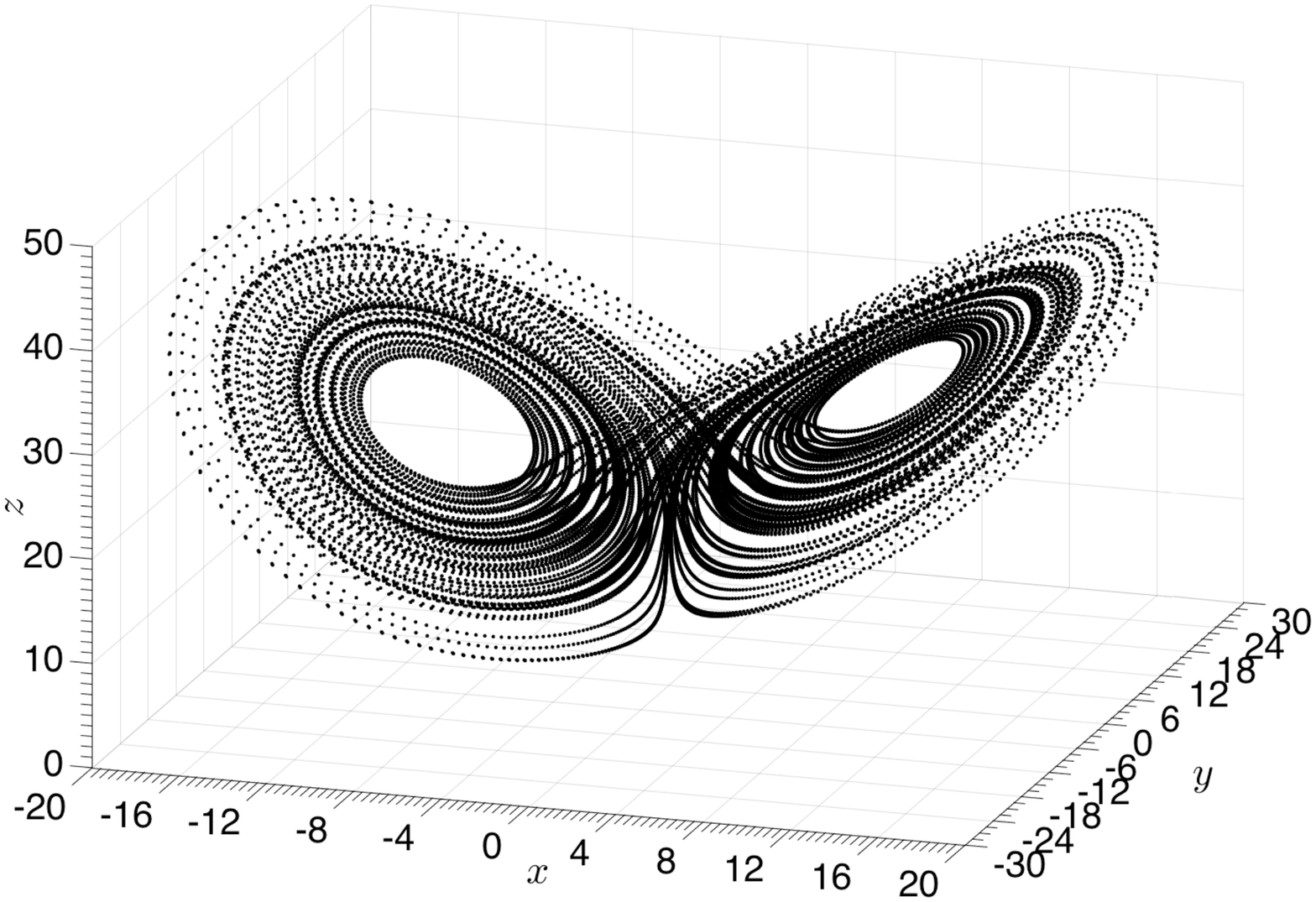}
\caption{Attractor of the Lorenz '63 system with indication of the cartesian grids used for constructing the partitions of its phase space. See text.}
\label{lorenzattr}
\end{figure}

As the box-counting dimension or capacity of the attractor of the model given in Eq. \ref{lorenz} is $d_0\sim2.05$, we expect that the number of boxes $B^j_k$, $k=1,\dots,N_B^j$ needed to cover the attractor decreases $N_B^j\propto 1/j^{d_0}$. We obtain a slightly lower exponent $\sim 1.9$, which is perfectly acceptable as we are far from the asymptotic regime where the scaling given by $d_0$ is realized. 

\begin{table}[ht]
\begin{center}
      \caption{Expectation value of the observables $x^2$, $y^2$, $z^2$,  and $z$ and their linear response with respect to the perturbation $\rho\rightarrow\rho+\epsilon$. The first row refers to the integration of the Lorenz model given in Eq, \ref{lorenz}. The other rows refer to the empirical discrete Markov chain constructed using boxes of different sizes. $N^j_{B}$ refers to the number of states. The linear response of the observables defined in Eq. \ref{sensitivity2} has been obtained using Eq. \ref{pertope3}. The derivative with respect to $\epsilon$ is estimated using finite differences with $\epsilon=0.1$. See text.}
          \label{tablelorenz}
    \begin{tabular}{|| l | c | c |  c | c || c | c | c | c  ||}
    \hline
     & $\langle x^2\rangle$ & $\langle y^2\rangle$ & $\langle z^2\rangle$ & $\langle z \rangle $ &$\delta [x^2]_1$ & $\delta [y^2]_1$ & $\delta [z^2]_1$&  $\delta [z]_1$ \\ \hline
    Lorenz '63 Model & 62.9 & 81.2 & 630.0 & 25.6 & 2.8 & 3.7 & 50.3 &1.01  \\ \hline
    MC, $j=1$, $N^j_{B}=770$& 63.2 & 82.0 & 630.5 & 23.6 & 2.9 & 3.8 & 50.3 &1.01   \\ \hline
   MC, $j=2$, $N^j_{B}=205$ & 64.3 & 84.2 & 632.2 & 23.6 & 3.0 & 3.5 & 49.7 &1.02 \\ \hline
   MC, $j=4$, $N^j_B=56$ & 71.3 & 84.8 & 637.5 & 23.5 & 2.9 & 3.9 & 50.1 &1.02 \\ \hline
    \hline
  \end{tabular}
\end{center}
\end{table}

For each value of $j$, the boxes $B^j_k$ define the discrete states $\phi^j_k$, $k=1,\dots,N_B^j$. By counting the number of times the trajectory is included in each state $\phi^j_k$ and normalizing we derive experimentally the asymptotic normalized occupancies $\bar{u}^j_k$. Instead, by tracking the transitions between the various discrete states, we construct the estimate of the stochastic transition matrix $\mathcal{M}^j_{p,q}$ describing the probability that the state $\phi^j_q$ makes a transition to the state $\phi^j_p$ in one time step. By finding the eigenvector corresponding to the unique unitary eigenvalue of $\mathcal{M}^j_{p,q}$, we find the invariant measure, which agrees up to very high precision with the empirical occupancy rate $\bar{u}_k$ computed from the trajectory. As a first step, we evaluate the expectation values of four meaningful observables given by $x^2$, $y^2$, $z^2$, and $z$, as obtained from the time integration of the Lorenz model and from its discrete representation in terms of Markov chain. Table \ref{tablelorenz} shows that the agreement is rather good even when extremely coarse resolution is used.    

We then show how to compute the response of the system to the perturbation due to the introduction of the vector field $\epsilon \mathbf{X}(\mathbf{x})$.  We keep in mind that when continuous time dynamics is considered, there is a very simple linear relation between the perturbation flow and the corresponding perturbation to the Perron-Frobenius operator, see Eqs. \ref{Mcont}-\ref{mcont2}. 

Therefore, we repeat the the steps described above for the $\epsilon-$perturbed flow (we choose $\epsilon=0.1$ in order to be on the safe side in terms of convergence), compute the new stochastic transition matrices $\mathcal{M}^{j,\epsilon}_{p,q}$, and derive the perturbation matrices $\epsilon m^j_{p,q}= \mathcal{M}^{j,\epsilon}_{p,q}-\mathcal{M}^{j}_{p,q}$. Once $m_{p,q}$ and $\mathcal{M}^{j}_{p,q}$ are known, we can use them to compute the  response of the systems at all orders of nonlinearity using Eqs. \ref{expressfinale} and \ref{pertope5}. One needs to note that because of the non-infinite  integration time considered, of the non-infinitesimal perturbation applied, and of the somewhat arbitrary choice of the boxes, it can  happen that the original and perturbed flow may be characterized by a different number of discrete states. We have observed such a difference only in the case $j=1$, involving one single extra state for the perturbed flow, with normalized relative occupancy ($\leq 10^{-6}$). This problem can be easily sorted out by imposing a cutoff and removing from the the discrete description all states with very low.   

As discussed above, one needs to test accurately the well-posedness and convergence of the expansion in order to be sure to obtain meaningful results. This is not our goal at this stage for such a preliminary numerical test of our results. Therefore, we limit ourselves to the less ambitious yet interesting goal of computing the linear response defined in Eq. \ref{sensitivity2} for the observables indicated above, using Eq. \ref{pertope3}. The results are reported in Table \ref{tablelorenz} and seem very encouraging. We have that the results are very stable with respect to changes in the resolution of the boxes, and agree to a high degree of precision with the results one obtains by empirically evaluating the sensitivity of the observables with respect to the introduction of the perturbation flow using two integrations, as well, in the case of the $z$ observable, with what reported in \cite{L09}. We note that the results are virtually unchanged if one uses instead of the high resolution time series with time step of 0.001 time units sparser observations corresponding to, \textit{e.g}. a time step of 0.01 time units. Obviously, using a time resolution lower by a factor of $s$ with respect to what considered here, one derives by tracking the transitions a stochastic transition matrix corresponding to the $s^{th}$ power of the one obtained at higher resolution. This does not affect the results as long as the sampling is much higher than the characteristic time scale of the system, which can be approximated in $\sim 1/\lambda_1\sim 1.1$ time units, where $\lambda_1$ is the positive Lyapunov exponent of the system. On longer time scales, instead, the stochastic matrix is quasi-degenerate, with all columns almost equal to the invariant measure 

\section{Conclusions}\label{comments}
Taking the point of view of finite state Markov systems, we have been able to construct a perturbation theory for studying the impact of small perturbations to the background dynamics. While previous approaches focus on the constructing a theory able to account for the effect of adding small perturbations to the baseline flow, we focus on computing  the change in the invariant measure and for the change in the expectation values of general observables (one problem being the adjoint of the other) occurring when the Markov transition matrix $\mathcal{M}\rightarrow\mathcal{M}+\epsilon m$. 

The perturbation term $\epsilon m$ has to be such that all the columns of the new stochastic matrix sum up to 1 and all entries are positive.  All of our findings are obtained with rather simple linear algebra manipulations and using basic properties of the stochastic matrices. We can express the response as a perturbation series or, after suitable resummation, using compact exact formulas. We are also able to assess the convergence properties of the response theory by defining a value $\epsilon^*_{max}$ such that if $|\epsilon|\leq \epsilon^*_{max}$ the perturbative expansion converges. We have that the stronger is the mixing of the unperturbed system, the larger is the value of $\epsilon_{max}$. These findings match well with previous results providing upper bounds to the sensitivity of stochastic matrices to perturbations.
%It is also extremely simple to change the point of view between studying the response of the invariant measure and looking in the expectation value of a generic observable, by computing the transpose of the resulting linear operators. 

Our results provide a direct algorithmic method for studying the response to perturbations for finite state Markov processes and have the advantage of allowing for an immediate and practical change of point of view between response theory seen in terms of changes of the invariant measure or in terms of changes in the expectation values of observables, by simply computing the transpose of the resulting finite dimensional linear operators. Our findings give  closed formulas for the linear and nonlinear response theory at all orders of perturbations through explicit matrix expressions that can be directly implemented in any coding language. %, plus providing bounds on the radius of convergence of the perturbative theory. In particular, we relate the convergence of the response theory with the rate of mixing of the unperturbed system, in agreement with previous results studying the sensitivity of the properties of stochastic matrices to perturbations.  

We can use our formulas to study the response to perturbations of finite state Markov processes constructed in order to have a simplified and treatable picture of a complex system. Given two different state spaces constructed using different finite partitions covering the attractor of the system, we cannot expect to obtain the same results for the change in the expectation value of a given observables.  The results might indeed be model dependent, but this is the obvious price one has to pay because of the subjective choice of the reduced state space. An assessment of the robustness of the obtained results is key to applying our methods in the context of reduced models. Nonetheless, the extremely unsophisticated numerical study reported here on the Lorenz'63 model is quite encouraging at this regard, even if test should be made on much higher dimensional models.

If the underlying dynamics is Axiom A (or Axiom A equivalent, as in the cases where the chaotic hypothesis applies), one can impose conditions such that the response operators constructed using finer and finer partitions converge to to the actual corresponding response operators constructed on the SRB measure. The conditions are stricter than what needed in order to have convergence of the unperturbed measure, the basic reason being that Ruelle response operators correspond to nontrivial observables. 

Our results can be thought as intermediate steps at finite precision leading to the correct response formulas in the limit.  One needs to add as a caveat that going from finite state to functional spaces is far from trivial and requires a high degree of mathematical precision, which is beyond the scopes of this paper. Nonetheless, the finite construction proposed here seems to somehow point at why some important mathematical issues emerge when the Perron-Frobenius operator formalism is considered (\textit{e.g.} selection of suitable norms for vectors and linear operators, definition of specific functional spaces for the observables).  

%Note that the result is general does not require performing the limit following a specific sequence of partitions as, \textit{e.g.}, the self-refining Markov partitions of the dynamics, which constitute a natural guess for a optimal set of finer and finer partitions in terms of convergence properties.

Interestingly, we can use the formulas obtained for finite state Markov processes to study the impact of perturbations to continuous time dynamical systems, after making a suitable identification between the considered transition matrices and the evolution operators for measures and observables. This operation is straightforward because there is a simple linear exact relation between the perturbation in the vector flow of the dynamical system and the perturbation in the Perron-Frobenius operator when infinitesimal time intervals are considered. As a result, we are able to derive in a very simple way previous formulas obtained studying the perturbations to the transfer operator as well as the original expressions proposed by Ruelle for the linear and higher order perturbations in the expectation values of observables. Using the results obtained in the finite state case, we propose a formula for the radius of expansion of the perturbative theory. 

One can envision that in the case the underlying dynamics is discrete, there is not such a one-to-one correspondence between perturbations to the vector field and perturbations to the Markov transition matrix. This can be easily checked when constructing the perturbed Perron-Frobenius operator resulting from adding a $\epsilon$ correction to the vector field, which results into changes in the Perron-Frobenius operator at all orders in $\epsilon$. Therefore, the perturbative expansion is different in the two cases. Agreement is instead found in the limit $\epsilon\rightarrow 0$, or, more practically, when we retain only the linear terms in $\epsilon$ perturbative expansion, \textit{i.e.} when aiming only at the linear response function. 

Future investigations will try, on the one side, to have a sharper mathematical look at the problem of going from finite to infinitely small partitions of the phase space, and, on the other side, to delve in the numerical study of the effectiveness and efficiency of the proposed tools. Apart from testing the results on specific finite state Markov systems, we will test how robust the proposed methods are when studying finite state Markov processes that have been empirically constructed from time series of observations or of numerical simulations of high-dimensional complex systems. One may be led to hoping that it could be possible to have an accurate representation of the response of a high dimensional system to perturbations by constructing a smart finite state model well suited to studying specific observables of interest. Of course, in order to deal with the curse of dimensionality, one would like to be able to go beyond the Ulam method and deal with finite partition of reduced phase spaces where projection is applied on many or even most dimensions. 

Our formulas may address the now long-standing problem of constructing suitable algorithms for studying the response of chaotic systems to perturbations.  It is extremely hard to construct an algorithm for computing the (linear) response theory directly on the flow, because serious problems emerge when considering the contributions coming from the unstable directions in the tangent space.  This might have great relevance for studying problems, like climate dynamics, where a direct construction of the response operator is especially challenging and slightly indirect methods have to be used \cite{LBHRPW14} and a lot of effort has been devoted to defining the so-called atmospheric regimes and predicting their response to forcings \cite{Corti1999}.

\subsubsection*{Acknowledgements}
VL wishes to thank: J. V\"ollmer for suggesting the author to look into finite state Markov processes; D. Ruelle and S. Vaienti for reading an earlier version of the manuscript; V. Baladi, G. Froyland,  T. Kuna, A. Tantet for many stimulating exchanges and for providing some extremely useful references and hints. VL acknowledges the support of the DFG-funded cluster of excellence CliSAP and of the FP7 ERC StG NAMASTE - Thermodynamics of the Climate System (Grant No. 257106). This paper is dedicated to Alexei Likhtman, a colleague who left us way too soon. 
\bibliographystyle{unsrt}

\end{document}